\documentclass[prl,preprint,superscriptaddress,showpacs]{revtex4}

\usepackage{amsmath}
\usepackage{amssymb}
\usepackage{graphicx}

\newcommand{\T}{${\mathcal T}\,$}
\newcommand{\Ti}{${\mathcal T}$}
\newcommand{\be}{\begin{equation}}
\newcommand{\ee}{  \end{equation}}
\newcommand{\ba}{\begin{eqnarray}}
\newcommand{\ea}{  \end{eqnarray}}

\begin{document}

\title{Induced Violation of Time-Reversal Invariance in the Regime of Weakly
Overlapping Resonances}

\author{B.~Dietz}
%\email{dietz@ikp.tu-darmstadt.de}
\affiliation{Institut f{\"u}r Kernphysik, Technische Universit{\"a}t
Darmstadt, D-64289 Darmstadt, Germany}

\author{T.~Friedrich}
%\email{friedrich@ikp.tu-darmstadt.de}
\affiliation{Institut f{\"u}r Kernphysik, Technische Universit{\"a}t
Darmstadt, D-64289 Darmstadt, Germany}
\affiliation{GSI Helmholtzzentrum f{\"u}r Schwerionenforschung GmbH, D-64291
Darmstadt, Germany}

\author{H.~L.~Harney}
%\email{hanns-ludwig.harney@mpi-hd.mpg.de}
\affiliation{Max-Planck-Institut f{\"u}r Kernphysik, D-69029 Heidelberg,
Germany}

\author{M.~Miski-Oglu}
%\email{maksim@ikp.tu-darmstadt.de}
\affiliation{Institut f{\"u}r Kernphysik, Technische Universit{\"a}t
Darmstadt, D-64289 Darmstadt, Germany}

\author{A.~Richter}
\email{richter@ikp.tu-darmstadt.de}
\affiliation{Institut f{\"u}r Kernphysik, Technische Universit{\"a}t
Darmstadt, D-64289 Darmstadt, Germany}
\affiliation{$\rm ECT^*$, Villa Tambosi, I-38100 Villazzano (Trento), Italy}

\author{F.~Sch{\"a}fer}
%\email{schaefer@ikp.tu-darmstadt.de}
\affiliation{Institut f{\"u}r Kernphysik, Technische Universit{\"a}t
Darmstadt, D-64289 Darmstadt, Germany}

\author{J.~Verbaarschot}
\affiliation{Department of Physics and Astronomy, SUNY at Stony Brook, NY
11794, USA}

\author{H.~A.~Weidenm{\"u}ller}
\affiliation{Max-Planck-Institut f{\"u}r Kernphysik, D-69029 Heidelberg,
Germany}

\date{\today}

\begin{abstract}
	We measure the complex scattering amplitudes of a flat
	microwave cavity (a ``chaotic billiard''). Time-reversal (\Ti)
	invariance is partially broken by a magnetized ferrite placed
	within the cavity. We extend the random-matrix approach to \T
	violation in scattering, determine the parameters from
	some properties of the scattering amplitudes, and then
	successfully predict others. Our work constitutes the most
	precise test of the theoretical approach to \T violation
	within the framework of random-matrix theory so far available.
\end{abstract}

\pacs{24.60.Ky, 05.45.Mt, 11.30.Er, 85.70.Ge}
% 24.60.Ky Fluctuation phenomena 
% 05.45.Mt Quantum chaos; semiclassical methods 
% 11.30.Er Charge conjugation, parity, time reversal, and other discrete symmetries
% 85.70.Ge Ferrite and garnet devices

\maketitle
We measure the effect of partial violation of time-reversal (\Ti)
invariance on the excitation functions of a flat microwave cavity
induced by a magnetized ferrite placed within the cavity. The
classical dynamics of a point particle moving within the cavity and
elastically reflected by the walls, is chaotic. The statistical
properties of the eigenvalues and eigenfunctions of the analogous
quantum system are, therefore, expected to follow random--matrix
predictions~\cite{Bohigas1984}. Random-matrix theory (RMT) provides a
universal description of generic properties of chaotic quantum
systems. In particular, RMT yields analytical expressions for
correlation functions of scattering amplitudes~\cite{Ver85} that can
be generalized to include \Ti violation. Although widely used (to
discover signatures of \Ti violation in compound-nucleus
reactions~\cite{Witsch1967} in the Ericson regime~\cite{Ericson}, to
describe electron transport through mesoscopic samples in the presence
of a magnetic field~\cite{Bergman}, and in ultrasound transmission in
rotational flows\cite{Rosny}), that generic model for \Ti violation
has, to the best of our knowledge, never been exposed to a detailed
experimental test. With our data we perform such a test.

Our aim is not a detailed dynamical modeling of the properties of the
cavity. With the exception of the average level density we determine
the parameters of the RMT expressions from fits to some of the data.
We then test the RMT approach by using it to predict other data, and
by subjecting our fits to a thorough statistical test. All of this is
in the spirit of a generic RMT approach since a dynamical calculation
of the relevant parameters is not possible for many systems. Such a
calculation works only for special chaotic quantum systems like some
cavities where the semiclassical approximation can be
used~\cite{Bluemel1998,Bluemel1990}. We use that approximation only to
determine the average level density, and to estimate the range of
validity of RMT in terms of the shortest periodic orbit.

Microwave cavities have been used before to study the effect of
\Ti-invariance violation on the
eigenvalues~\cite{So1995,Stoffregen1995,Hul2004} and on the
eigenfunctions~\cite{Wu1998}. Here we study fluctuations of the
scattering amplitudes versus microwave frequency. For our cavity the
average resonance spacing $d$ is of the order of the resonance width
$\Gamma$, and we work in the regime of weakly overlapping resonances.

\begin{figure}[hb]
	\centering \includegraphics[width=8cm]{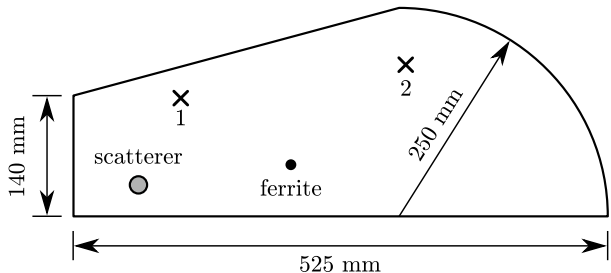} \caption{The
	tilted stadium billiard (schematic). The two antennas $1$, $2$
	connect the resonator to the VNA. The ferrite is fixed, the
	scatterer can be moved freely.}  \label{fig:tilted_stadium}
\end{figure}

\emph{Experiment.} The flat copper microwave resonator has the shape of a
tilted stadium~\cite{Primack1994} (see Fig.~\ref{fig:tilted_stadium})
and a height of 5~mm. The excitation frequency $f$ ranges from 1 to
25~GHz. In that range, only one vertical mode of the electric field
strength is excited. The Helmholtz equation for the tilted stadium is
then mathematically equivalent to the Schr{\"o}dinger equation of a
two-dimensional chaotic quantum billiard~\cite{Stoeckmann1990}. An
Agilent PNA-L~N5230A vector network analyzer (VNA) coupled rf energy
via one of two antennas labeled $1$ and $2$ into the resonator and
determined magnitude and phase of the transmitted (reflected) signal
at the other (same) antenna in relation to the input signal and, thus,
the elements $S_{a b}(f)$ with $a, b = 1, 2$ of the complex-valued
$2\times2$ scattering matrix $S(f)$. Distorting effects of the
connecting coaxial cables were removed by calibration. We measured the
elements of $S(f)$ in the frequency range 1--25~GHz at a resolution of
100~kHz. To improve the statistical significance of the data set, an
additional scatterer (an iron disc of 20~mm diameter) was placed
within the cavity. It could be freely moved and allowed the
measurement of statistically independent spectra, so-called
``realizations''.

\begin{figure}[hb]
	\centering \includegraphics[width=8cm]{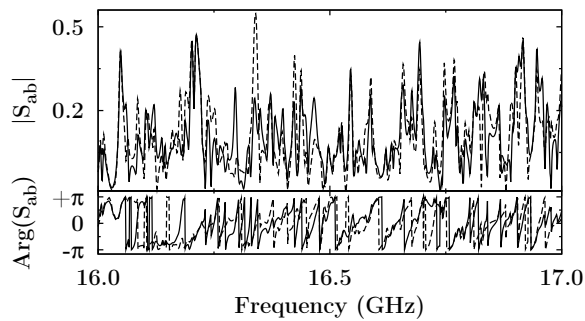}
	\caption{Transmission spectra for $B = 190~{\rm mT}$ in the
	range 16--17~GHz. The amplitudes and phases of $S_{12}$
	(solid) and $S_{21}$ (dashed) are seen to differ.}
	\label{fig:fluct_reci}
\end{figure}

Time-reversal invariance is violated~\cite{Schaefer2007} by a ferrite
cylinder ($4\pi\, M_{\rm S} = 1859~{\rm Oe}$, $\Delta H = 17.5~{\rm
Oe}$, courtesy of AFT Materials GmbH, Backnang, Germany) of 4~mm
diameter and 5~mm height. The cylinder was placed inside the resonator
and magnetized by an external magnetic field $B$. The field was
provided by two NdFeB magnets (cylindrical shape, 20~mm diameter and
10~mm height) attached from the outside to the billiard. Field
strengths of up to 360~mT could be attained. Here we focus on the
results at $B = 190~{\rm mT}$ as there the effects are most clearly
visible. The spins within the ferrite precess collectively with their
Larmor frequency about the external field. The rf magnetic fields of
the resonator modes are, in general, elliptically polarized and couple
to the spins of the ferrite. The coupling depends on the rotational
direction of the rf field. An interchange of input and output channels
changes the rotational direction and thus the coupling of the
resonator modes to the ferrite. Figure~\ref{fig:fluct_reci}
demonstrates that reciprocity, defined by $S_{12}(f) = S_{21}(f)$ and
implied by \T invariance, is violated.

\begin{figure}[ht]
	\centering \includegraphics[width=8cm]{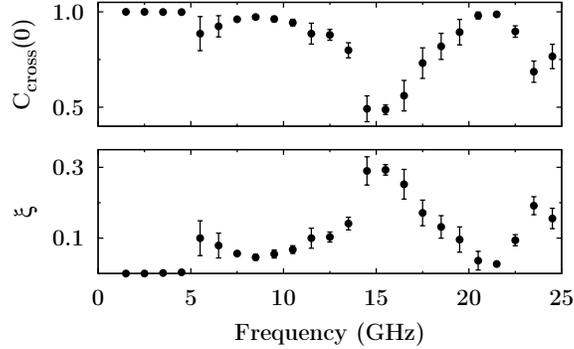}
	\caption{Experimentally determined values of $C_{\rm
	cross}(0)$ (upper panel) from Eq.~(\ref{eqn:cross}) and the
	parameter $\xi$ for \Ti-invariance violation deduced from
	these (lower panel) with the help of Eq.~(\ref{eqn:Cab}). The
	error bars indicate the r.m.s.\ variation of $C_{\rm
	cross}(0)$ over the 6 realizations.}  \label{fig:alpha}
\end{figure}

As a measure of the strength of \Ti-invariance violation, we define
the cross-correlation coefficient $C_{\rm cross}(\epsilon=0)$ where
\begin{equation}
	C_{\rm cross}(\epsilon) = \frac{\mathfrak{Re}\left(\langle
	S_{12}(f) \, S_{21}^\ast(f+\epsilon)\rangle\right)} 
	{\sqrt{ \langle |S_{12}(f)|^2\rangle\, \langle |S_{21}(f)|^2
	\rangle}} \, .
	\label{eqn:cross}
\end{equation}
If \T invariance holds, we have $C_{\rm cross}(0) = 1$ while for
complete breaking of \T invariance $S_{12}$ and $S_{21}$ are
uncorrelated and thus $C_{\rm cross}(0) = 0$. The average $\langle\,
\cdot\, \rangle$ over the data is taken in frequency windows of width
$1~$GHz and over 6 realizations, i.e., positions of the additional
scatterer. The upper panel of Figure~\ref{fig:alpha} shows $C_{\rm
cross}(0)$ for the different frequency windows. The cross-correlation
coefficient is seen to depend strongly on $f$ although complete
violation of \T invariance is never attained. At 5--7~GHz the Larmor
frequency of the ferrite matches the rf frequency, and the
ferromagnetic resonance directly results in $C_{\rm cross} (0) \approx
0.8$.  Around 15~GHz the effects of \Ti-invariance violation are
strongest, $C_{\rm cross}(0) \approx 0.5$. A third minimum is observed
at about 24~GHz. The connection of the latter two minima to the
properties of the ferrite is not clear.

\emph{Analysis.} We analyze the data with a scattering approach developed in
the context of compound-nucleus reactions~\cite{Mahaux1969}. The
scattering matrix for the scattering from antenna $b$ to antenna $a$
with $a,b=1,2$ is written as
\begin{equation}
	S_{ab}(f) = \delta_{ab} - 2\pi\, i\, \big( W^\dagger\,
	(f-H^{\rm eff})^{-1}\, W \big)_{a b} \, .
	\label{eqn:Sab}
\end{equation}
The matrix $W_{\mu a}$ is rectangular and describes the coupling of
the $N$ resonant states $\mu$ in the cavity with the antennas $a = 1,
2$. We assume that \Ti-invariance violation is due to the ferrite
only. Then $W_{\mu a}$ is real. The resonances in the cavity are
modeled by $H^{\rm eff} = H - i\pi\,\tilde{W} \tilde{W}^\dag$. Here
$H$ is the Hamiltonian of the closed resonator. The elements of the
real matrix $\tilde{W}_{\mu c}$ are equal to those of $W_{\mu c}$ for
$c=1,2$. As done successfully before~\cite{Uzy,Friedrich2008}, Ohmic
absorption of the microwaves in the walls of the cavity and the
ferrite is mimicked~\cite{Brouwer1997} by additional fictitious weakly
coupled channels $c$. The classical dynamics of a point particle
within the tilted stadium billiard is
chaotic. Therefore~\cite{Bohigas1984}, we model $H$ by an ensemble of
random matrices. The $N$-dimensional Hamiltonian matrix $H$ of the
system (the cavity) is written as the sum of two parts
\cite{Pandey1981,Pandey1983,Altland1993}, $H = H^s + i(\pi \xi /
\sqrt{N}) H^a$. The real, symmetric, and \Ti-invariant matrix $H^s$ is
taken from the Gaussian Orthogonal Ensemble (GOE) while the real,
antisymmetric matrix $H^a$ with Gaussian-distributed matrix elements
models the \Ti-invariance breaking part of $H$. For $\pi \xi /
\sqrt{N} = 1$ the Hamiltonian $H$ belongs to the Gaussian Unitary Ensemble (GUE)
describing systems with complete \T breaking. However, for $N
\to \infty$, \T invariance is significantly broken already when the
dimensionless parameter $\xi$ is close to unity~\cite{footnote}. In
the same limit $C_{\rm cross}(0)$ in Eq.~(\ref{eqn:cross}) can be
expressed analytically in terms of a threefold integral involving the
parameter $\xi$. For the derivation we extended the method of
Ref.~\cite{Pluhar1995} where the ensemble average of $\vert
S_{ab}\vert^2$ was computed as function of the parameter $\xi$. The
cross-correlation coefficient $C_{\rm cross}(0)$ is obtained by
setting $\epsilon=0$, $\sigma =-1$, and $a,b=1,2$ in the function
\begin{eqnarray}
	&&F^\sigma_{ab}(\epsilon\vert T_a,T_b,\tau_{\rm abs},\xi) =
	\frac{1}{8} \int_0^\infty{\rm d}x_1\int_0^\infty{\rm
	d}x_2\,\int_0^1 {\rm d}x\, \nonumber\\
	& \times&\frac{\mu(x,x_1,x_2)}{\mathcal{F}}\cdot
	\exp\left(-\frac{i\pi\epsilon}{d}(x_1+x_2+ 2x)\right)
	\nonumber\\
	&\times&\prod_c \frac{1-T_c\, x}{\sqrt{(1+T_cx_1) (1+T_c\,x_2)} }
	\Big[ \Big\{ J_{ab}(x,x_1,x_2)
	\nonumber\\
	&\times&\left[ \mathcal{F} \mathcal{E}_+ + (\lambda_2^2 -
	\lambda_1^2) \mathcal{E}_- + 4\mathfrak{t}\mathcal{R}
	(\lambda_2^2
	\mathcal{E}_-+\mathcal{F}(\mathcal{E}_+-1))\right] \nonumber\\
	&+&\sigma\cdot 2(1 - \delta_{ab})T_aT_b
	\big[\mathcal{E}_-K_{ab}(\lambda,\lambda_1,\lambda_2\vert
	T_a,T_b,\xi ) \nonumber\\
	&+&\left(\mathcal{E}_+-\frac{\mathcal{E}_-}{t\mathcal{F}}\right)
	L_{ab}(\lambda,\lambda_1,\lambda_2\vert T_a,T_b,\xi )\big]
	\Big\} \cdot\exp\left( -2\mathfrak{t}\mathcal{G}_- \right) 
	\nonumber\\
	&+& \left(\lambda_1 \leftrightarrow \lambda_2\right)\Big], 
	\label{eqn:Cab}                  
\end{eqnarray}
with the notations
\begin{eqnarray}
	\mathfrak{t}&=&\pi^2\,\xi^2,\quad \mathcal{R} = 4(x + x_1)(x +
	x_2), \nonumber\\
	\mathcal{U}&=&2\sqrt{x_1 (1+x_1)x_2(1+x_2)},\quad \mathcal{F} = 
	4x(1-x), \nonumber \\
	\mathcal{E}_\pm&=&1 \pm \exp(-2\mathfrak{t}\mathcal{F}),\quad
	\lambda=1 - 2x,\nonumber\\
	\lambda_i&=&\sqrt{ (1+x_1)\,(1+x_2)+x_1x_2 -(-1)^i\,
	\mathcal{U}},\nonumber\\ \mathcal{G}_i&=&\lambda_i^2-1,
	i=1,2\, .  \label{eqn:defs}
\end{eqnarray}
The integration measure $\mu(x,x_1,x_2)$ and the function
$J_{ab}(x,x_1,x_2)$ are given explicitly in
Ref.~\cite{Ver85}, while the functions
$K_{ab}(\lambda,\lambda_1,\lambda_2\vert T_a,T_b,\xi )$ and
$L_{ab}(\lambda,\lambda_1,\lambda_2\vert T_a,T_b,\xi )$ can be read
off Eq.~(2) of Ref.~\cite{Gerland}. We checked our analytic results by
numerical RMT simulations. The parameters of Eq.~(\ref{eqn:Cab}) for
$\epsilon=0$ are $\xi$, the transmission coefficients $T_a = 1 -
|\langle S_{a a} \rangle |^2$ for $a = 1, 2$, and the sum $\tau_{\rm
abs}$ of 300 transmission coefficients that model the Ohmic
losses~\cite{Brouwer1997,Friedrich2008}.

\begin{figure}[hb]
	\centering \includegraphics[width=8cm]{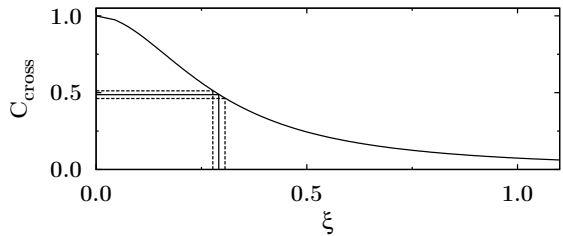}
	\caption{Dependence of the cross-correlation coefficient
	$C_{\rm cross}(0)$ on the parameter $\xi$ as predicted by the
	random-matrix model for partial violation of \T
	invariance. Also shown is how $C_{\rm cross}(0)=0.49(3)$
	translates into $\xi=0.29(2)$.}  \label{fig:cross_vs_lambda}
\end{figure}

For a typical set $T_1, T_2, \tau_{\rm abs}$,
Fig.~\ref{fig:cross_vs_lambda} shows $C_{\rm cross}(0)$ versus
$\xi$. Within the frequency range 1--25 GHz, $C_{\rm cross}(0)$
depends very weakly on $T_1, T_2, \tau_{\rm abs}$, and
Fig.~\ref{fig:cross_vs_lambda} can be taken to be universal.
For each data point shown in the upper panel of Fig.~\ref{fig:alpha}
the corresponding value of $\xi$ was read off
Fig.~\ref{fig:cross_vs_lambda} and the result is shown as function of
$f$ in the lower panel of Fig.~\ref{fig:alpha}.
Figure~\ref{fig:alpha} shows that in the interval from 1 to 25 GHz,
the ratio of the average resonance width $\Gamma$ to the average
resonance spacing $d$ varies from $\Gamma/d \approx 0.01$ to $\Gamma/d
\approx 1.2$ while the strength $\xi$ of \T breaking varies from zero
to $0.3$. Numerical calculations show that for $\xi =0.3$ the spectral
fluctuations of the Hamiltonian $H$ for the closed resonator defined
below Eq.~(\ref{eqn:Sab}) almost coincide with those of the
GUE~\cite{Bohigas1995}. We also found that for $\xi=0.4$ they do not differ 
significantly from those presented in Ref.~\cite{So1995}, where the conclusion 
was drawn, that complete \T breaking is achieved. However, even for $\xi =0.4$
the value of $C_{\rm cross}(0)$ is still far from zero. This shows that
$C_{\rm cross}(0)$ is a particularly suitable measure of the strength
$\xi$ of \T violation.

\begin{figure}[hb]
	\centering \includegraphics[width=8cm]{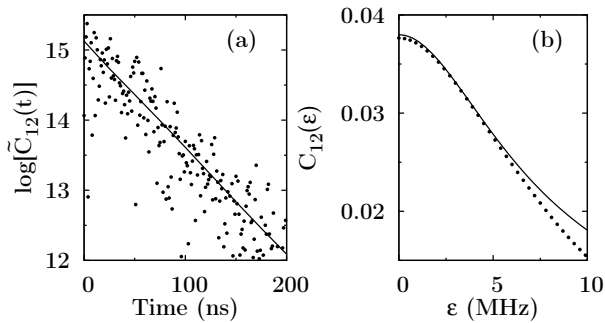}
	\caption{Autocorrelation function for $S_{1 2}$ in the range
	of 16--17~GHz and at $B = 190~{\rm mT}$. In the time domain
	(a) the data (dots) scatter around the theoretical fit (solid)
	for $T_1=0.37$, $T_2=0.41$, $\tau_{\rm abs}=2.9$ and $\xi =
	0.25$. Transforming the results back into frequency domain (b)
	confirms the good agreement between data and theory. We
	observe that neighboring data points in (b) are correlated,
	whereas those in (a) are not.}  \label{fig:correl}
\end{figure}

\emph{Autocorrelation function.} Since $C_{\rm cross}(0)$ depends only weakly
on the values of $T_1, T_2$ and $\tau_{\rm abs}$, we used the
autocorrelation function $C_{ab}(\epsilon)$ for a more precise
determination of these parameters, especially of $\tau_{\rm abs}$.
The function
\begin{equation}
	C_{ab}(\epsilon) = \langle S_{ab}(f)\,
	S_{ab}^\ast(f+\epsilon) \rangle - | \langle S_{ab}(f)
	\rangle |^2 \label{eqn:auto}
\end{equation}
was calculated analytically with the method of Ref.~\cite{Pluhar1995}
as a function of $T_1, T_2, \tau_{\rm abs}, \xi$, and $d$ and is
obtained from Eq.~(\ref{eqn:Cab}) by setting $\sigma = +1$. It
interpolates between the well-known results for orthogonal
symmetry~\cite{Ver85} (full \T invariance) and for unitary
symmetry~\cite{Fyodorov2005} (complete violation of \T invariance).
The mean level spacing $d$ was computed from the Weyl formula. The
Fourier transform of the function $C_{ab}(\epsilon)$ was then fitted to the data as in
Ref.~\cite{Friedrich2008}. As starting points we used the values of
$T_1$ and $T_2$ obtained from the measured values of $S_{a a}(f)$ and
of $\xi$ determined from $C_{\rm cross}(0)$. For each of the $6$
realizations the spectra of $S_{a b}(f)$ were divided into intervals
$\Delta f$ of 1~GHz length. In each interval the Fourier transform
$\tilde{C}_{ab}(t_k)$ of the autocorrelation function~(\ref{eqn:auto})
was calculated for values of $t_k$ between $5~{\rm ns}$ and $200~{\rm
ns}$. The lower limit is determined by the length of the shortest
periodic orbit in the classical billiard; for smaller values of $t_k$
the Fourier coefficients are nongeneric~\cite{Bluemel1990}. At $t_k
\approx 200~{\rm ns}$ the values of $\tilde{C}_{ab}(t_k)$ have decayed
over more than three orders of magnitude, and noise limits the
analysis. The time resolution was $1/\Delta f = 1~{\rm ns}$. We
measured four excitation functions $S_{a b}(f)$ taking $(a, b) = (1, 1),
(1, 2), (2, 1), (2, 2$ yielding a total of
$4800$ Fourier coefficients for each interval. For $f >$ 10 GHz the
fitted values for $T_1$ and $T_2$ differ by not more than 7~\% from
the initial ones. (For smaller $f$ the intervals of 1 GHz width
comprise only few resonances). The spread of the data is large, see
the left panel of Fig.~\ref{fig:correl}. Going to the time domain is
useful since the $S_{a b}(f)$ are correlated for neighboring $f$
whereas the correlations are removed in the ratios of the experimental
and the fitted values for $\tilde{C}_{ab}(t_k)$. The latter are
stationary and fluctuate about unity.  Thus the statistical analysis
is much simplified.

For each realization the parameters $\tau_{\rm abs}$ and $\xi$ were
obtained by fitting the analytical expression for
$\tilde{C}_{ab}(t_k)$ to the experimental results. The values of $\xi$
determined from these fits agree with the ones found from the
cross-correlation coefficient. To reduce the spread we combined the
data from all realizations within a fixed frequency interval. The
result was analyzed with the help of a goodness-of-fit (GOF) test (see
Ref.~\cite{Friedrich2008}) that distinguishes between full, partial,
and no violation of \T invariance. We defined a confidence limit such
that the GOF test erroneously rejects a valid theoretical description
of the data with a probability of 10~\%. With this confidence limit
the test rejects the fitted expressions for $\tilde{C}_{ab}(t_k)$ in
only 1 out of the 24 available frequency windows or in 4.2~\% of the
tests. Thus, the RMT model correctly describes the fluctuations of the
$S$-matrix for partial violation of \T invariance in the regimes of
isolated and weakly overlapping resonances.

\begin{figure}[ht]
	\centering \includegraphics[width=8cm]{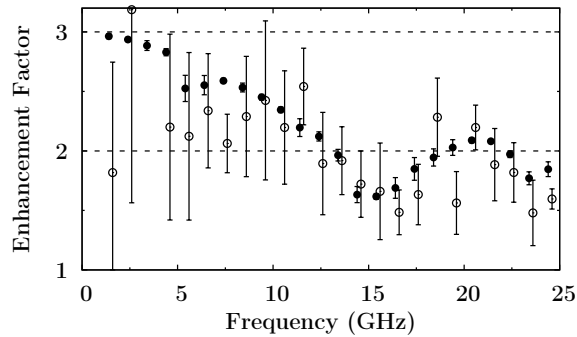}
	\caption{Comparison of elastic enhancement factors. For the
	evaluation of $\mathcal{W}$ the autocorrelation coefficients
	$C_{ab}(0)$ were determined either directly from the data
	(open circles) or from the analytic result for partial
	violation of \T invariance (filled circles) with $\xi$
	determined from the cross-correlation coefficient. The error
	bars indicate the variations within the 6 realizations. The
	dashed horizontal lines indicate the limits of $\mathcal{W}$
	for \Ti-invariant systems in the regime of isolated
	($\mathcal{W} = 3$) or overlapping ($\mathcal{W} = 2$)
	resonances.}  \label{fig:enh_190mT}
\end{figure}

\emph{Elastic enhancement factor.} As a second test of the theory we use 
the values of $\xi$ obtained from the cross-correlation coefficients
(see Fig.~\ref{fig:alpha}) and the parameters $T_a, T_b, \tau_{\rm
abs}$ resulting from the fit of $\tilde{C}_{ab}(t_k)$ to predict the
values of the elastic enhancement factor $\mathcal{W} = \bigg( \langle
|S^{\rm fl}_{1 1}|^2\rangle \ \langle |S^{\rm fl}_{2 2}|^2 \rangle
\bigg)^{1/2} / \ \langle |S^{\rm fl}_{1 2}|^2 \rangle$ with $S^{\rm
fl}_{\rm ab} = S_{\rm ab} - \langle S_{\rm ab} \rangle$ as a function
of $f$. We use that $\mathcal{W} = \sqrt{C_{11}(0)\, C_{22}(0)} /
C_{12}(0)$, see Eq.~(\ref{eqn:auto}). For \Ti-invariant systems, the
elastic enhancement factor decreases from $\mathcal{W} = 3$ for
isolated resonances with many weakly coupled open channels to
$\mathcal{W} = 2$ for strongly overlapping resonances ($\Gamma \gg
d$). The corresponding values for complete violation of \T invariance
are $\mathcal{W} = 2$ and $\mathcal{W} = 1$,
respectively~\cite{Savin2006}. Figure~\ref{fig:enh_190mT} compares the
analytic results for the enhancement factor $\mathcal{W}$ (filled
circles) to the data (open circles). For small $f$ (where $\Gamma / d
\ll 1$ and $\xi \approx 0$) the experimental results differ from the
prediction $\mathcal{W} = 3$. Here only few resonances contribute and
the errors of the experimental values for $\mathcal{W}$ are large.
Moreover $\mathcal{W}$ is determined from only a single value $C_{a
b}(0)$ of the measured autocorrelation function while the analytic
result is based on a fit of the complete autocorrelation function.
As $f$ increases so does $\Gamma / d$, and $\mathcal{W}$ takes values
well below 3. At frequencies where $\xi$ is largest $\mathcal{W}$
drops below the value $2$ as predicted, a situation that cannot arise
for \Ti-invariant systems. The overall agreement between both data
sets above $\approx 10~{\rm GHz}$ corroborates the confidence in the
values of $\xi$ deduced from the cross-correlation coefficients.

\emph{Summary.} We have investigated partial violation of \T invariance with
the help of a magnetized ferrite placed inside a flat microwave
resonator (a chaotic billiard) with two antennas. We measured
reflection and transmission amplitudes in the regime of isolated and
weakly overlapping resonances in the frequency range from $1$ to $25$
GHz and determined the cross-correlation function, the autocorrelation
functions, and the elastic enhancement factor from the data. The
results were used as a test of random-matrix theory for scattering
processes with partial \T violation. That theory yields analytic
expressions for all three observables. The parameters of the theory
($T_1, T_2, \tau_{\rm abs}$ and the parameter $\xi$ for \T violation)
were partly obtained directly from the data but improved values
resulted from fits to the autocorrelation function. We find that $0
\leq \xi \leq 0.3$. The validity of the theory was tested in two ways.
(i) A goodness-of-fit test of the Fourier coefficients of the
scattering matrix in frequency intervals of $1$ GHz width yielded
excellent agreement. (ii) The elastic enhancement factor predicted
from the fitted values of the parameters shows overall agreement with
the data for frequencies above $10$ GHz where the experimental errors
are small. We conclude that the random-matrix description of
$S$-matrix fluctuations with partially broken \T invariance is in
excellent agreement with the data.

\begin{acknowledgments}
F.~S.\ is grateful for the financial support from the Deutsche Telekom
Foundation. This work was supported by the DFG within SFB~634.
\end{acknowledgments}


\begin{thebibliography}{10}

\bibitem{Bohigas1984}
O. Bohigas, M. J. Giannoni, and C. Schmit, Phys. Rev. Lett.
{\bf 52}, 1 (1984).

\bibitem{Ver85}
J.~J.~M. Verbaarschot, H.~A. Weidenm\"uller, and M.~R. Zirnbauer,
Phys. Rep. {\bf 129}, 367 (1985).

\bibitem{Witsch1967}
W. von Witsch, A. Richter, and P. von Brentano, Phys. Rev. Lett.
{\bf 19}, 524 (1967); E. Blanke {\it et~al.}, ibid. {\bf 51}, 355 (1983).

\bibitem{Ericson}T. Ericson, Phys. Rev. Lett. {\bf 5}, 430 (1960).

\bibitem{Bergman}
G. Bergman, Phys. Rep. {\bf 107}, 1 (1984).

\bibitem{Rosny}
J. Rosny {\it et~al.} Phys. Lett. {\bf 95}, 074301 (2005). 

\bibitem{Bluemel1998}
R. Bl\"umel and U. Smilansky, Phys. Rev. Lett. {\bf 60}, 477 (1988).

\bibitem{Bluemel1990}
R. Bl\"umel and U. Smilansky, Phys. Rev. Lett. {\bf 64}, 241 (1990).

\bibitem{So1995}
P. So, S.~M. Anlage, E. Ott, and R.~N. Oerter, Phys. Rev. Lett. {\bf
74}, 2662 (1995).

\bibitem{Stoffregen1995}
U. Stoffregen {\it et~al.}, Phys. Rev. Lett. {\bf 74},  2666  (1995).

\bibitem{Hul2004}
O. Hul {\it et~al.}, Phys. Rev. E {\bf 69},  056205  (2004).

\bibitem{Wu1998}
D.~H. Wu, J.~S.~A. Bridgewater, A. Gokirmak, and S.~M. Anlage,
Phys. Rev. Lett. {\bf 81}, 2890 (1998).

\bibitem{Primack1994}
H. Primack and U. Smilansky, J. Phys. A {\bf 27}, 4439  (1994).

\bibitem{Stoeckmann1990}
H.-J. St\"ockmann and J. Stein, Phys. Rev. Lett. {\bf 64}, 2215
(1990).

\bibitem{Schaefer2007}
B. Dietz {\it et~al.}, Phys. Rev. Lett. {\bf 98}, 074103  (2007).

\bibitem{Mahaux1969}
C. Mahaux and H.~A. Weidenm\"uller, {\em Shell-Model Approach to Nuclear
Reactions} (North-Holland Publ. Co., Amsterdam, 1969).

\bibitem{Uzy} 
C.~H.~Lewenkopf and A. M{\"u}ller, Phys. Rev. A {\bf 45}, 2635;
R.~Sch{\"a}fer, T.~Gorin, T.~H.~Seligman, and H.-J-~St{\"o}ckmann,
J. Phys. A {\bf 36}, 3289 (2003).

\bibitem{Friedrich2008}
B. Dietz {\it et~al.}, Phys. Rev. E {\bf 78},  055204(R)  (2008).

\bibitem{Brouwer1997}
P.~W. Brouwer and C.~W.~J. Beenakker, Phys. Rev. B {\bf 55}, 4695
(1997).

\bibitem{Pandey1981}
A. Pandey, Ann. Phys. (N.Y.) {\bf 134}, 110 (1981).

\bibitem{Pandey1983}
A. Pandey and M. L. Mehta, Comm. Math. Phys. {\bf 87}, 449 (1983).

\bibitem{Altland1993}
A.~Altland, S.~Iida, and K.~B.~Efetov, J. Phys. A {\bf 26}, 3545 (1993).

\bibitem{footnote} 
Time-reversal invariance is significantly broken for $\pi\xi /
\sqrt{N} \simeq d/v$. Here, $d = v\pi / \sqrt{N}$ denotes the average
level spacing and $v^2$ the variance of the off-diagonal matrix
elements of $H^{\rm s }$, $H^{\rm a}$. 

\bibitem{Pluhar1995}
Z. Pluha\v{r} {\it et~al.}, Ann. Phys. {\bf 243}, 1 (1995).

\bibitem{Gerland}
U. Gerland and H. A. Weidenm\"uller, Europhys. Lett. {\bf 35}, 701
(1996).

\bibitem{Bohigas1995} 
O.~Bohigas, M.~J. Giannoni, A.~M. Ozorio de Almeida, and C. Schmit, 
Nonlinearity {\bf 8}, 203 (1995).

\bibitem{Fyodorov2005}
Y.~V. Fyodorov, D.~V. Savin, and H.-J. Sommers, J. Phys. A:
Math. Gen. {\bf 38}, 10731 (2005).

\bibitem{Savin2006}
D.~V. Savin, Y.~V. Fyodorov, and H.-J. Sommers, Acta Phys. Pol. A {\bf
109}, 53 (2006). 

\end{thebibliography}
\end{document}